\documentclass[twocolumn,aps,prl,showpacs]{revtex4}

\usepackage{graphicx}

\begin{document}

\title{Stable two-dimensional ferromagnets made of regular single-layered lattices of
single-molecule nanomagnets on substrates}

\author{Rui Zheng and Bang-Gui Liu }
\affiliation{Institute of Physics, Chinese Academy of Sciences,
Beijing 100190, China\\ Beijing National Laboratory for Condensed
Matter Physics, Beijing 100190, China}

\date{\today}

\begin{abstract}
We propose that stable two-dimensional (2D) ferromagnets can be made
of regular single-layered lattices of single-molecule nanomagnets
with enough uniaxial magnetic anisotropy on appropriate substrates
by controlling the inter-nanomagnet magnetic interaction. Our Monte
Carlo simulated results show that such ideal 2D ferromagnets are
thermodynamically stable when the anisotropy is strong enough. If
the anisotropy energy equals 80 K, approximately that of the
Mn$_{12}$, the $T_c$ varies from zero to 15 K depending on different
inter-nanomagnet coupling constants. Such stable spin systems,
experimentally accessible, should be promising for information
applications.
\end{abstract}


\maketitle


Single-molecule (SM) nanomagnets attract huge interest because
they show quantum tunnelling, interference, and coherence effects
at ultra-low temperatures and have various potential applications,
such as magnetic information storage and quantum
computation\cite{1,2,3,4,5,6,7,8}. An SM nanomagnet usually has
large spin $S$ and strong uniaxial anisotropy, and there is an
easy axis for spin orientation\cite{8}. Hence it must overcome an
effective energy barrier $U_{\rm eff}$ to achieve a spin reversal
at low enough temperature. For example, the famous Mn$_{12}$ SM
nanomagnet
([Mn$_{12}$O$_{12}$(O$_{2}$CCH$_{2}$Br)$_{16}$(H$_{2}$O)$_{4}$]$\cdot$4CH$_{2}$Cl$_{2}$)
has $S=10$ and its spin reversal energy barrier $U_{\rm eff}$ is
upto 74 K\cite{8,9}. The record-breaking value of $U_{\rm eff}$ is
86.4 K in the case of [Mn$^{I\! I\!
I}_6$O$_2$(Et-sao)$_6$(O$_2$CPh(Me)$_2$)$_2$$\cdot$(EtOH)$_6$],
which has $S=12$\cite{10}.

SM magnets can be made to form single isolated magnets, dimers,
and other clusters\cite{1,2,3,4,5}. Through chemical methods, they
can be synthesized in the form of three-dimensional (3D)
crystalline structures and two-dimensional (2D)
networks\cite{9,10,a3d1,a3d2,a2d}. The basic building blocks in
these structures are single magnets. The effective inter-magnet
magnetic interactions usually are antiferromagnetic (AF) and
cannot be easily controlled. Thus, it is quite difficult to make
stable ferromagnets from such systems of SM magnets. Fortunately,
single layers of SM nanomagnets have been recently achieved on
gold and silicon surfaces\cite{11,12,13}. This breakthrough makes
us believe that stable ideal 2D ferromagnets can be made from
regular 2D lattices of SM nanomagnets, although Mermin-Wagner
theorem allows no finite-temperature phase transition in 2D
isotropic Heisenberg spin systems\cite{14,15}.

Here we show that stable 2D ferromagnets can be made of regular
single-layered lattices of SM nanomagnets with enough uniaxial
anisotropy on appropriate substrates by controlling the
inter-nanomagnet magnetic interaction. The spin of the nanomagnet
is described as a large spin. The easy axis is perpendicular to
the 2D spin lattice. We describe the 2D ferromagnets by a 2D
quantum Heisenberg model with the uniaxial anisotropy and
relatively weak ferromagnetic (FM) inter-spin coupling. This model
is reasonable and liable because the dipolar spin interaction
almost is reduced to a purely AF one and the inter-spin coupling
can be controlled by choosing right substrates and manipulating
the inter-nanomagnet distances\cite{16}. We reduce the quantum
spin operators to classical variables because both the spin value
and the anisotropy energy are large enough\cite{8,9,10}, and then
use Monte Carlo method to simulate the many-body physical
properties with various parameters\cite{17,18}. Our results prove
that stable 2D ferromagnets can be made of such SM nanomagnets for
experimentally-accessible parameters. More detailed results will
be presented in the following.

SM nanomagnets such as the Mn$_{12}$ have large spins and their
inter-spin exchange interactions usually are very weak compared to
their anisotropy energies\cite{8,9,10,16}. When put on appropriate
surfaces, such SM nanomagnets can have the same uniaxial
anisotropy energy and adjustable inter-spin
interactions\cite{11,12,13}. Thus, the 2D ferromagnets can be
described by the Hamiltonian
\begin{equation}
H=-K\sum_i(S^z_i)^2-\sum_{ij}J_{ij}\vec{S}_i\cdot\vec{S}_j
\end{equation}
where $K$ describes the uniaxial anisotropy with the easy axis
($z$ direction) perpendicular to the 2D lattice, the $\vec{S}_i$
is the spin operator at site $i$ in the lattice, and $J_{ij}$ is
the inter-spin coupling constant between $\vec{S}_i$ and
$\vec{S}_j$. This perpendicular anisotropy and the 2D lattice
structure make the dipolar spin interaction almost reduced to an
AF one, and thus $J_{ij}$ can varies from AF to FM interaction.
Here we are interested mainly in the FM case. The latter summation
is over spin pairs between $i$ and $j$. Because the spin $S$ is
very large and we are interested in the equilibrium properties, we
can reasonably treat the quantum spin operator $\vec{S}_i$ as a
classical vector $S\vec{s}_i$, where $\vec{s}_i$ is a unit vector.

We investigate the equilibrium properties of this model using Monte
Carlo (MC) method\cite{17,18}. Both Metropolis and cluster
algorithms are used\cite{19,20}. We take an $L\times L$  square
lattice and limit $J_{ij}$ to nonzero $J$ ($>0$) only for the
nearest spin pairs without losing main physics. As usual, we use a
periodic boundary condition. For convenience, we use a special unit
that the Boltzmann constant $k_B$ is set to 1. We use the parameter
$D=K/J$ to describe the strength of the anisotropy energy relative
to the exchange constant. As we prove, the system is already in
equilibrium after 100,000 MC steps (MCS). The average magnetization
$M$, the fourth-order Binder's cumulant $U_4$, the specific heat
$C$, and the magnetic susceptibility  $\chi$ are calculated by
averaging over 50,000 MCS after thermodynamical equilibrium is
reached\cite{17,18}. Ten copies of such average values are used to
calculate the final results.

\begin{figure}[!htbp]
\begin{center}
\includegraphics[width=.45\textwidth]{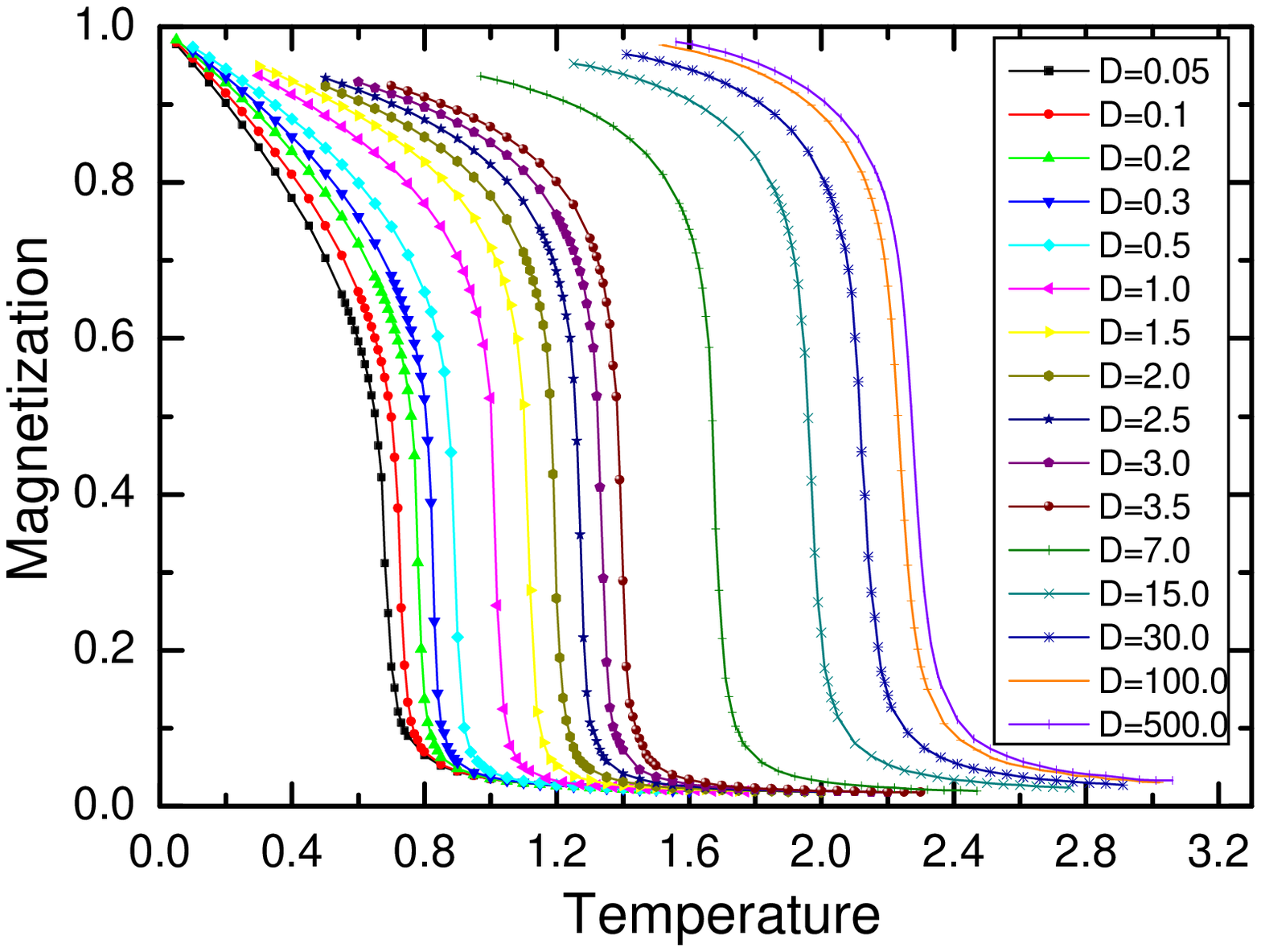}
\caption{(color online). Normalized magnetizations $M$ as functions
of temperature $T$ for given anisotropy parameters from $D$=0.05 to
500.} \label{fig1}
\end{center}
\end{figure}

In Figure 1 we present average magnetizations $M$ as functions of
temperature $T$ for 16 $D$ values from 0.05 to 500. We have $M=1$
at $T=0$ because the magnetization $M$ is normalized to 1. The
magnetization $M$ for all the higher temperatures increases with
$D$. There is a spinodal point in the specific magnetization curve
for a given $D$ value. When $D$ becomes larger, the spinodal point
shifts rightward to higher temperature, which means that the
phase-transition temperature $T_c$ increases with increasing $D$.
In order to determine $T_c$ more accurately, we actually use the
unique intersection point of the various $U_4-T$ curves of
different $L$ values. This method is the best approach to obtain
accurate phase-transition temperature\cite{17,18}. The calculated
$T_c$ as function of $D$ is presented in Figure 2. In order to
emphasize the $T_c$ of large $D$ values, we use logarithmic scale
in the main plot and linear scale in the inset. Our results show
that $T_c$ increases with $D$, converging to 2.269, the Ising
limit\cite{21,22}, when $D$ is infinite. For $D=0$ we obtain
$T_c=0$, which is consistent with established conclusion that
there is no finite-temperature phase-transition for 2D isotropic
model\cite{14,15}. A nonzero $D$ is necessary to a finite $T_c$.

\begin{figure}[!htbp]
\begin{center}
\includegraphics[width=.42\textwidth]{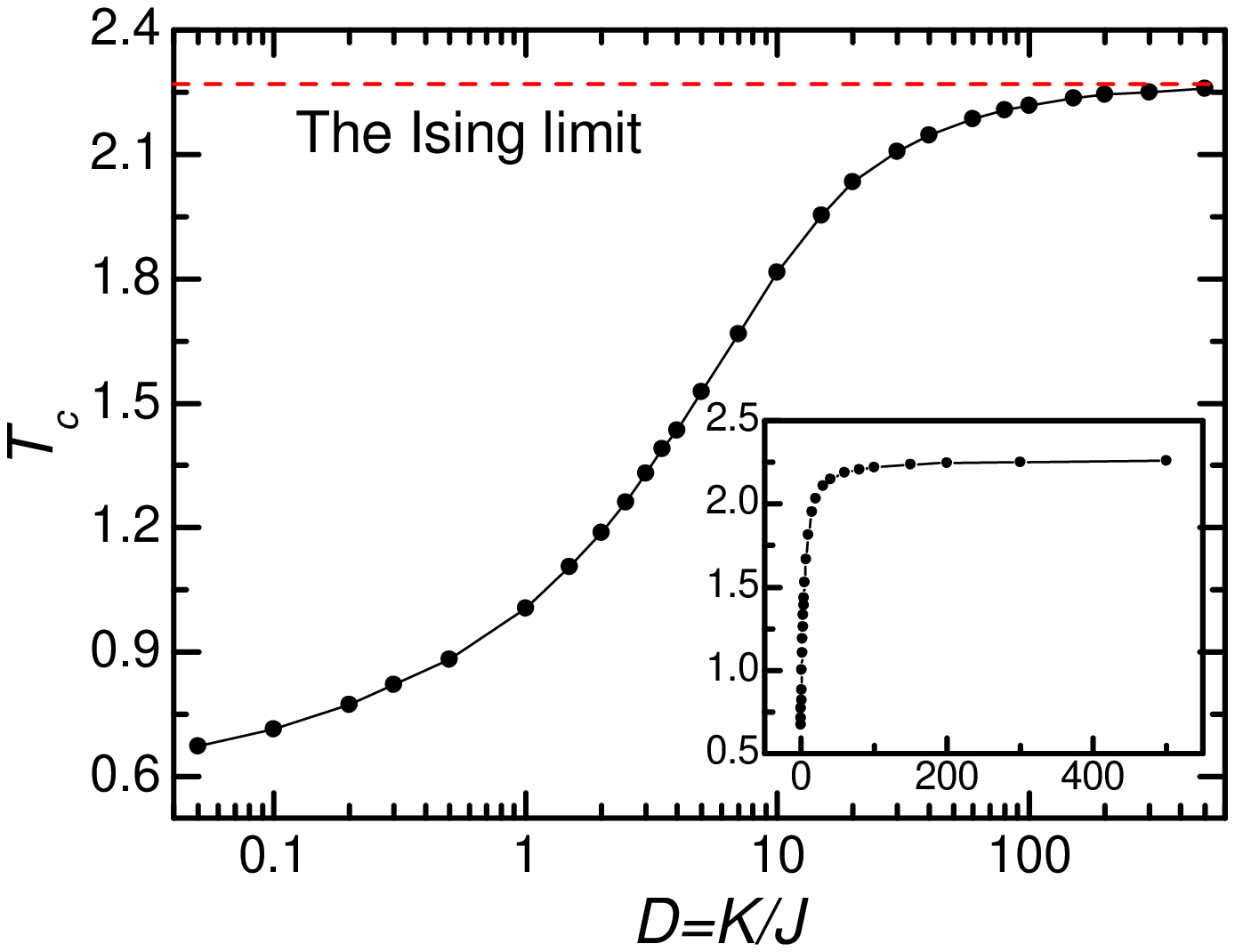}
\caption{(color online). The $D$ dependence of Curie temperature
$T_c$ with $D$ from 0.05 to 500. The dash line indicates Tc=2.269,
the Ising limit. The inset shows $T_c$ as a function of $D$ in
linear scale.} \label{fig2}
\end{center}
\end{figure}

We describe the deviation of the normalized spin vector from the
easy axis by the parameter $d_m$ defined as
\begin{equation}d_m=1-\langle (s^z)^2\rangle /\langle
(\vec{s}^2)\rangle,\end{equation} where
$\vec{s}=(1/L^2)\sum\vec{s}_i$ and $\langle A\rangle$ is the
average value of $A$. We present $d_m$($T$) for various $D$ values
in Figure 3. It is clear that $d_m$ is zero or tiny at low
temperature. When $D$ approaches to zero, we obtain $d_m\leq 2/3$.
For a given $D$, $d_m$ increases with $T$, reaching a maximum
$d_{mc}$ at $T_c$. $d_{mc}$ decreases with increasing $D$, being
already less than 0.02 for $D=30$. For $D\geq 100$, $d_m$ is
actually zero in the temperature window in Figure 3. Indeed, when
$D$ is very large, the Hamiltonian is similar to the 2D Ising
model, but it is equivalent to the 2D Ising model only when $D$ is
infinite.

\begin{figure}[!htbp]
\begin{center}
\includegraphics[width=.45\textwidth]{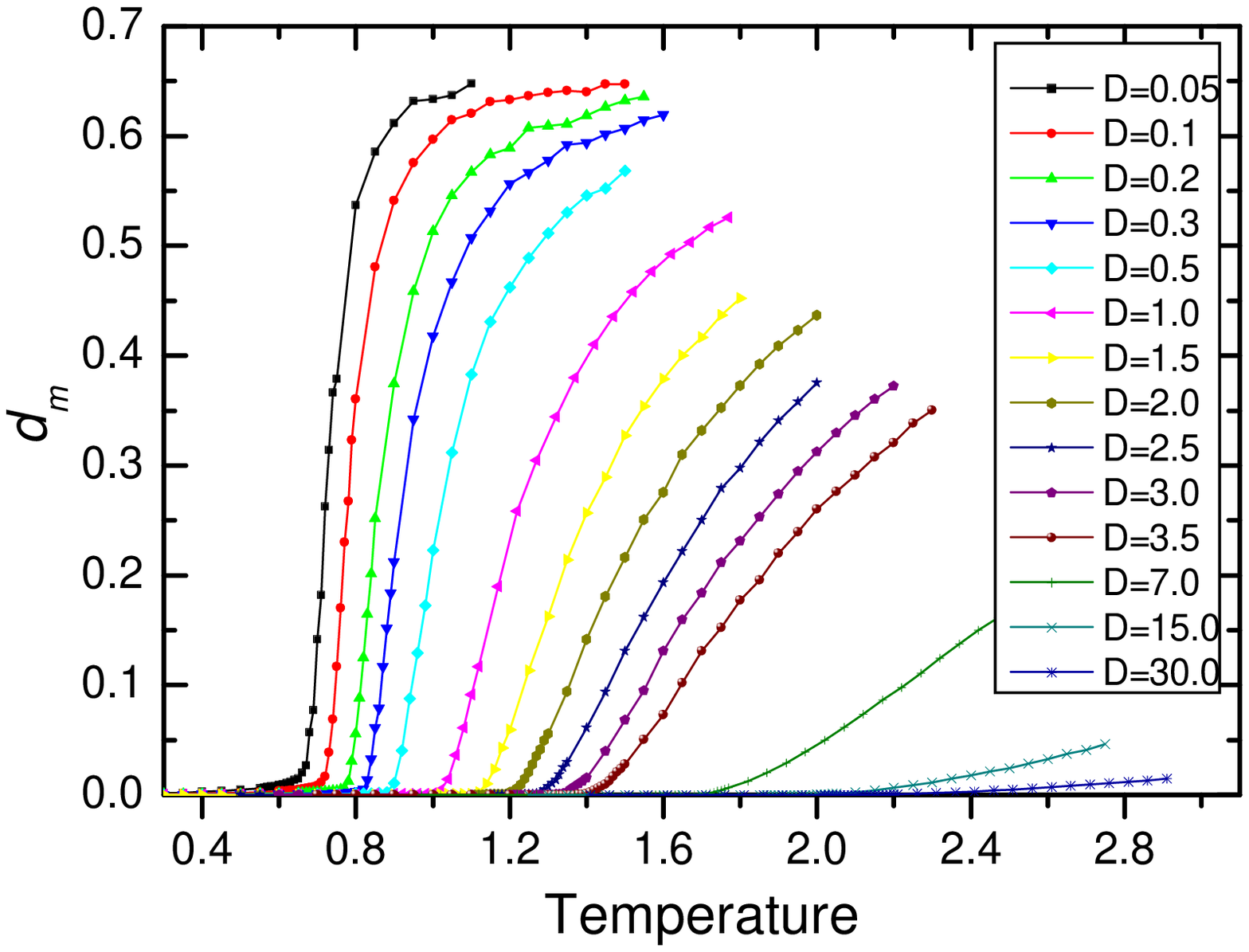}
\caption{(color online). The temperature dependence of the deviation
$d_m$ of the spin vector from the easy axis for various $D$ values
(from 0.05 to 30.0).} \label{fig3}
\end{center}
\end{figure}

It is established that the inter-spin interaction can be changed
experimentally and even adjusted intentionally for special
purposes\cite{16}. This means that $J$ can be changed actually from
a few K to zero. Considering that the parameter $K$ in Eq. (1) can
be nearly 80 K\cite{10}, we conclude that the anisotropy parameter
$D$ can be 10 or much larger. Accordingly, $T_c$ is $1.8\sim
2.269J$. With $K=80$ K, $T_c$ equals 2.2 K if $J=1$ K, and 14.4 K if
$J=8$ K. Because the 2D ferromagnets are in the regime of large $D$,
the spin deviation $d_m$ is very small, smaller than 0.001, when $T$
is at $T_c$, and can be considered zero when $T$ is substantially
below $T_c$. Therefore, the spin is nearly in the Ising limit in
well-ordered FM states and such spin systems should be stable enough
to carry information.

In summary, we propose that stable 2D ferromagnets can be made of
regular single-layered lattices of SM nanomagnets with enough
uniaxial magnetic anisotropy on appropriate substrates by
controlling the inter-nanomagnet magnetic interaction. The quantum
spin operator for them can be naturally reduced to a classical
spin variable because the spin value and the anisotropy energy are
so large. Our Monte Carlo simulated results show that such ideal
2D ferromagnets are stable because the deviation of the spins from
the easy axis is smaller than 0.1\% in the ferromagnetic state
when the anisotropy is strong enough. If the anisotropy energy
equals 80 K, approximately that of the famous Mn$_{12}$, the $T_c$
varies from zero to $\sim$15 K according to the different
inter-spin coupling constants. Our 2D model is reasonable and
reliable because the dipolar inter-spin interaction almost can be
reduced to an AF one due to the perpendicular easy axis and the 2D
spin lattice. Therefore, such stable spin systems should be
promising for information applications.

\begin{acknowledgments}
This work is supported  by Nature Science Foundation of China (Grant
Nos. 10874232 and 10774180), by the Chinese Academy of Sciences
(Grant No. KJCX2.YW.W09-5), and by Chinese Department of Science and
Technology (Grant No. 2005CB623602).
\end{acknowledgments}


\begin{references}

\bibitem{1}  W. Wernsdorfer, N. Aliaga-Alcalde, D. N. Hendrickson, and
G. Christou, Nature 416, 406 (2002).

\bibitem{2}  W. Wernsdorfer and R. Sessoli, Science 284, 133 (1999).

\bibitem{3}  S. Hill, R. S. Edwards, N. Aliaga-Alcalde, G. Christou,
Science 302, 1015 (2003).

\bibitem{4} C. Schlegel, J. van Slageren, M. Manoli, E. K. Brechin, and
M. Dressel, Phys. Rev. Lett. 101, 147203 (2008).

\bibitem{5}  S. Bertaina, S. Gambarelli, T. Mitra, B. Tsukerblat, A.
Mueller and B. Barbara, Nature 453, 203 (2008).

\bibitem{6}  M. N. Leuenberger and D. Loss, Nature 410, 789 (2001).

\bibitem{7}  L. Bogani and W. Wernsdorfer, Nat. Mater. 7, 179 (2008).

\bibitem{8}  D. Gatteschi, R. Sessoli, and J. Villain, Molecular
nanomagnets, Oxford University Press 2006.

\bibitem{9}   N. E. Chakov, S.-C. Lee, A. G. Harter, P. L. Kuhns, A. P.
Reyes, S. O. Hill, N. S. Dalal, W. Wernsdorfer, K. A. Abboud, and
G. Christou, J. Am. Chem. Soc. 128, 6975 (2006). 

\bibitem{10}  C. J. Milios, A. Vinslava, W. Wernsdorfer, S. Moggach, S.
Parsons, S. P. Perlepes, G. Christou, and E. K. Brechin, J. Am.
Chem. Soc. 129, 2754 (2007). 

\bibitem{a3d1} H. Miyasaka, K. Nakata, K. Sugiura, and R. Clerac,
Angew. Chem., Int. Ed. 43, 707 (2004). 

\bibitem{a3d2} C. Coulon, R. Clerac, W. Wernsdorfer, T. Clin, and
H. Miyasaka, Phys. Rev. Lett. 102, 167204 (2009). 

\bibitem{a2d} H. Miyasaka {\it et al}, J. Am. Chem. Soc. 128, 3770
(2006). 

\bibitem{11}  B. Fleury, L. Catala, V. Huc, C. David, W. Z. Zhong, P.
Jegou, L. Baraton, S. Palacin, P. Albouyd, and T. Mallah, Chem.
Commun. 2005, 2020 (2005).

\bibitem{12} M. Cavallini, J. Gomez-Segura, D. Ruiz-Molina, M. Massi, C.
Albonetti, C. Rovira, J. Veciana, and F. Biscarini, Angew. Chem.
Int. Ed.  44, 888 (2005).

\bibitem{13}  A. Naitabdi, J. P. Bucher, P. Gerbier, P. Rabu, and M.
Drillon, Adv. Mater. 17, 1612 (2005).

\bibitem{14}  N. D. Mermin and H. Wagner, Phys. Rev. Lett. 17, 1133
(1966).

\bibitem{15}  L. D. Landau and E. M. Lifshitz, Statistical Physics Vol.
5, 482 (Pergmon, London, 1959).

\bibitem{16}  W. Wernsdorfer, S.
Bhaduri, A. Vinslava, and G. Christou, Phys. Rev. B 72, 214429
(2005).

\bibitem{17} K. Binder and D. W. Heermann, Monte Carlo Simulation in
Statistical Physics, Springer, Berlin, 2002.

\bibitem{18} K. Binder, Z. Phys. 43, 119 (1981).

\bibitem{19} N. Metropolis, A. W. Rosenbluth, M. N. Rosenbluth, A. M.
Teller, and E. Teller, J. Chem. Phys. 21, 1087 (1953).

\bibitem{20}  R. H. Swendsen and J.-S. Wang, Phys. Rev. Lett. 58, 86
(1987).

\bibitem{21} L. Onsager, Phys. Rev. 65, 117 (1944).

\bibitem{22} E. Ising, Z. Phys. 31, 253 (1925).
\end{references}
\end{document}